\documentclass[12pt]{iopart}
\expandafter\let\csname equation*\endcsname\undefined
\expandafter\let\csname endequation*\endcsname\undefined

\usepackage{amsmath}
\usepackage{graphicx, subcaption, hyperref}
\usepackage[dvipsnames]{xcolor}
\usepackage{ulem}

\begin{document}

%\title[Spatio-temporally non-local closure for multiscale modeling]{Machine learning hidden variables for spatio-temporally non-local physics in multiscale simulations}

%[Just a suggestion:] 
\title[Machine learning of hidden variables]{Machine learning of hidden variables in multiscale fluid simulation}

\author{Archis S. Joglekar}
\address{Ergodic LLC, San Francisco, CA 94117, USA \footnote{Also affiliated with the Department of Nuclear Engineering and Radiological Sciences, University of Michigan, Ann Arbor, MI 48109, USA}}
\ead{archis@ergodic.io}
\author{Alexander G. R. Thomas}
\address{Gerard Mourou Center for Ultrafast Optical Sciences, University of Michigan, Ann Arbor, MI 48109, USA}

\vspace{10pt}
\begin{indented}
\item[]\today
\end{indented}

\begin{abstract}
Solving fluid dynamics equations often requires the use of closure relations that account for missing microphysics. For example, when solving equations related to fluid dynamics for systems with a large Reynolds number, sub-grid effects become important and a turbulence closure is required, and in systems with a large Knudsen number, kinetic effects become important and a kinetic closure is required. By adding an equation governing the growth and transport of the quantity requiring the closure relation, it becomes possible to capture microphysics through the introduction of  ``hidden variables'' that are non-local in space and time. The behavior of the ``hidden variables'' in response to the fluid conditions can be learned from a higher fidelity or ab-initio model that contains all the microphysics. In our study, a partial differential equation simulator that is end-to-end differentiable is used to train judiciously placed neural networks against ground-truth simulations. We show that this method enables an Euler equation based approach to reproduce non-linear, large Knudsen number plasma physics that can otherwise only be modeled using Boltzmann-like equation simulators such as Vlasov or Particle-In-Cell modeling.
\end{abstract}

%
% Uncomment for keywords
\vspace{2pc}
\noindent{\it Keywords}: neural operators, plasma physics, kinetics, machine learning, differentiable physics
%
% Uncomment for Submitted to journal title message
%\submitto{\JPA}
%
% Uncomment if a separate title page is required
%\maketitle
% 
% For two-column output uncomment the next line and choose [10pt] rather than [12pt] in the \documentclass declaration
%\ioptwocol
%

\section{Introduction}
Machine learning for partial differential equations (PDEs) has received much attention with an emphasis towards computational fluid dynamics \cite{brenner_perspective_2019, duraisamy_turbulence_2019, brunton2020machine, vinuesa_enhancing_2022}.  Fluid equations describe conservation laws for bulk properties of matter --- such as mass, energy, average flow velocity --- under the continuum assumption. The equation set usually requires a closure, which involves additional information that represents some ``microphysics'' that is not captured by the continuum model. For example, because of turbulence on small scales, or granularity because of the nature of the media being made up of particles / molecules / cells etc. an equation may be introduced that represents the relationship between bulk properties, such as a constituent relation (material response to a force) or equation of state (a  thermodynamic equation relating bulk properties) While analytic closure models based on equilbrium states have existed at least since the introduction of hydrodynamic equations, %by Bernoulli}, 
the abundance of data and of data-generation mechanisms has prompted a re-examination of the existing methods in light of data-driven models that provide closure for the coupled system of equations. This approach is particularly relevant in systems where physics occurs on length- and time- scales smaller than that which is resolved by the simulations. %For example, the application of machine learning to model turbulence has a growing body of research (see references within refs. \cite{vinuesa_enhancing_2022, shankar_differentiable_2023}). 

Similarly, there are other systems where the fluid approximation itself breaks down because of effects involving an extended phase-space of states; for example, including momentum states where physical effects from the interaction between molecules or particles of a certain velocity become important. In these cases, increasing the resolution of the simulated domain arbitrarily does not recover the missing \emph{kinetic} physics.  Capturing kinetic physics using a first-principles approach requires a fully kinetic description such as solving the Boltzmann equation. Such effects may be included in a fluid model through constituent relations etc. that depend on the local conditions that may be derived or learned somehow from a kinetic model or experimental data etc. An example of this is the introduction of modified transport coefficients for nonlocal heatflow in fusion plasma \cite{Brodrick_POP_2017}.  More recently, the application of machine learning to model turbulence has a growing body of research (see references within refs. \cite{vinuesa_enhancing_2022, shankar_differentiable_2023}).

The issue with any relation derived from local bulk fluid conditions is that it retains no ``memory'' of microphysics effects from other locations; the effect is localized in space and time. But it is often the case that nonlocality is important. For example in the case of kinetic flows, fast particle streams will affect transport far, in both time and space, from a region that generates them.

In this article, we expand the method of machine learned closure to learn variables that themselves obey a PDE that describes the space-time variation of the microphysics. We refer to these as ``hidden variables'' as they do not describe bulk variables of the fluid system of equations but instead represent some physics that is ``hidden'' from the fluid equations. The original term ``hidden variables'' was introduced by Bohm to try to explain quantum behavior in terms of some unobservable underlying variables \cite{Bohm_PR_1952}. Here, the meaning is slightly different as although our ``hidden variable'' indeed describes physics that is unobservable (within the fluid description), the variable itself is not thought to be fundamental but just an ad-hoc reduced fidelity approximation of the fundamental behavior which can be learned from an ab-initio model. We specifically look at the dynamics of Landau damping of excited plasma wavepackets in the presence of  trapped hot electrons, which is described by the higher fidelity Vlasov-Poisson equations. Our fluid model using a Landau closure, similar to that in Ref.~\cite{hammett_fluid_1990}, that is modified to include a hidden variable representing the effect of the hot electron population is able to replicate the fully kinetic behavior. The structure of paper is as follows. Section \ref{sec1} discusses the inclusion of kinetic effects in a new differentiable fluid simulator using a ``hidden variable'' approach. Section \ref{sec2} introduces the specific problem to address of Landau damping of excited plasma wavepackets and the basic equations to be solved. Section \ref{sec3} introduces the new ``hidden variable'' and its evolution. Section \ref{sec4} presents the results and comparison with full Vlasov simulation. Finally, we conclude in Section \ref{sec5}.

\section{Inclusion of kinetic effects in a fluid code}\label{sec1}
The Boltzmann equation is a PDE that governs the evolution of the distribution function of particles in phase space and is given by 
\begin{align}
    \frac{\partial f}{\partial t} + \frac{\mathbf{p}}{m} \cdot \frac{\partial f}{\partial \mathbf{x}} + \mathbf{F} \cdot \frac{\partial f}{\partial \mathbf{p}} = \left(\frac{\partial f}{\partial t}\right)_{\text{coll}}, \label{eq:bman}
\end{align}
where $f=f(t, \mathbf{x}, \mathbf{p})$ is the distribution function in $(\mathbf{x}, \mathbf{p})$ phase space, and $\mathbf{F}$ are the forces, and the right-hand-side describes collisional effects. Velocity-space moments of eq. \ref{eq:bman} give an infinite chain of equations for the conservation of particle density, momentum, and energy. With appropriate closure conditions, these form the Euler equations in the inviscid case, or the Navier-Stokes equations in the viscous regime, depending on the treatment of the collisional term. If the particles are charged, this moment-driven approach gives the plasma fluid equations.

In all cases, the reduction of a three configuration and three velocity phase-space down to a three-dimensional configuration space averages over the details contained in velocity-space. This approximation is justified in systems where the Knudsen number is small such that the particles reach local thermal equilibrium (LTE) and velocity-space effects, called \emph{kinetic effects}, are negligible. The Knudsen number is defined by the ratio between the mean-free-path of the particles and the size of the system. There are many important systems where the Knudsen number is not small, and the thermal equilibrium approximation does not hold. For example, in micro- or nano- scale flows where the system size is small, as well as rarified gases such as the high-altitude atmosphere that is experienced by spacecraft where the mean-free-path is large. 

In plasmas, kinetic effects are important in the Scrape-Off Layer (SOL) in magnetic fusion plasmas \cite{Brodrick2017}, in the low-density gas-fill in inertial fusion plasmas \cite{Brodrick2017}, and in laser- and plasma-based accelerator experiments where the plasma density is low and the physics is weakly-collisional. In such settings, it becomes important to retain the influence of kinetic effects in order to perform accurate modeling of the system dynamics. Another example is in plasma based particle acceleration \cite{Albert_NJP_2021}, in particular the phenomena of wavebreaking or trajectory crossing in plasma waves \cite{Thomas_PRE_2016}. Fluid simulations of plasma accelerators have the benefit of having smooth distributions compared to typical particle-in-cell simulations, but cannot model crucial electron trapping processes because of phase space mixing.

%While similar to what is discussed here, kinetic effects occur in an orthogonal space entirely, and cannot be encompassed by simulating at a finer scale. Therefore, it is not enough to replace a closure term with a machine-learned term and one must capture the remaining physics using different methods. 

Previous attempts at the inclusion of kinetic effects into plasma fluid modeling have leveraged analytical methods \cite{hammett_fluid_1990, dimits_fast_2014, hunana_new_2018, fan_kinetic_2022}. Modeling kinetic physics using data-driven methods has either been attempted towards benchmarking their performance against known analytical methods \cite{cheng_data-driven_2022} or is restrictive in the geometries to which it can be applied \cite{lamy_modeling_2022}. In this work, we develop a machine-learned model that is applied to model a phenomenon with no known analytical representation. It is trained in a small geometry and successfully translated to a much larger domain.

%(a) We train a model for the growth rate of a hidden variable that serves to suppress the damping of a plasma wave. A naive local model (d) does not reproduce finite length effects that we see from Boltzmann-like simulations (d). Including this effect in a multiphysics simulation tool requires the implementation of a spatiotemporally non-local model (c) that can be accomplished by solving for the growth and transport of a hidden variable (e).

To accurately model the desired kinetic effects in this work, we find that a model that is non-local in space \emph{and} time is necessary. In other words, we find that a model that has \emph{memory} is required. To accomplish this, we use an additional transport equation for a hidden variable with learned coefficients for the source term. Because we leverage a hidden variable that can persist over time, it enables the reproduction of the kinetic effect that requires non-locality in time. Since our hidden variable can also be transported, it can be leveraged to reproduce non-local behavior in space.  We train the model in a reduced, idealized system against numerical solutions of the collisionless Boltzmann equation. Because our model is a self-consistent system of coupled PDEs with carefully constructed asymptotic behavior, we show that it not only extends to larger systems but can also be used to model phenomena that the neural networks have never previously seen. 

\begin{figure}[ht]
    \centering
    \includegraphics[width=\textwidth]{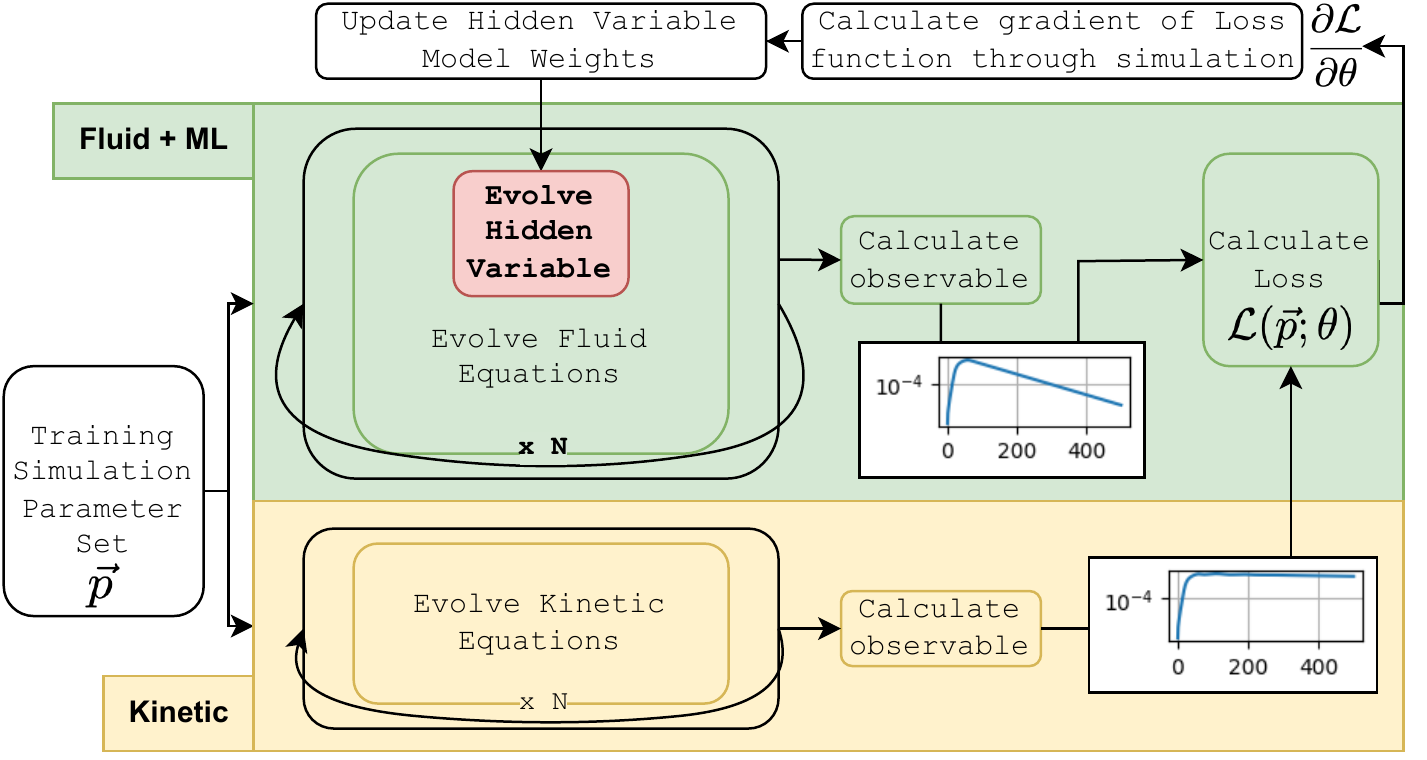}
    \caption{A single step in the training loop is illustrated. A batch of physical parameters is chosen. (Bottom) Those parameters are then fed to a kinetic simulator and an observable is calculated. (Top) The same parameters are fed to the differentiable fluid simulator that includes the machine learned hidden variable. This system is evolved in time over the same duration as the kinetic simulation. The observables from each simulation are used for calculating the loss function. The loss, and more importantly, the gradient of the loss with respect to the weights of the function approximator is computed. That gradient is used by an optimization algorithm to update the weights of the hidden variable model. The updated weights are used in the next batch of simulations. In practice, we precalculate the reference simulations.}
    \label{fig:all}
\end{figure}
In order to maximize the likelihood of developing models that are stable over long time-scales, it is important to develop models either using symbolic or sparse regression techniques \cite{rudy_data-driven_2017, alves_data-driven_2022, kaptanoglu_sparse_2023} or via differentiable simulators \cite{bar-sinai_learning_2019, kochkov_machine_2021}. The former allow the user to craft and analyze symbolic formulations and ensure their stability. In the most general case, however, formulating a library of terms that may enable accurate modeling is a challenging task, and using a function approximator such as a neural network provides a method by which to circumvent that challenge. Training neural network models in-situ with simulations, rather than only using a single time-step, helps ensure the stability of the learned representation.  This paradigm is visualized in fig. \ref{fig:all}.

To increase the likelihood of learning stable models, we choose to develop a differentiable fluid simulator for plasma physics called \textsc{ADEPT}\footnote{Automatic-Differentiation-Enabled Plasma Transport}. While differentiable simulators have been used to construct models for computational fluid dynamics \cite{bar-sinai_learning_2019, kochkov_machine_2021, shankar_differentiable_2023}, and for physics discovery in plasma kinetics \cite{joglekar_unsupervised_2022}, to our knowledge, this is the first differentiable simulator of computational plasma fluid dynamics. The code is publicly available along with the dataset introduced in this work.

%This work shows that this method enables the inclusion of kinetic effects into fluid modeling but it is important to stress that this method can be applied towards sub-grid physics as well. There has been some preliminary work in using this approach towards sub-grid modeling. In ref. \cite{diffburg}, the authors use a differentiable simulator to benchmark various turbulence closures, where a few of the benchmarked methods use an approach of modeling a hidden variable. 

%However, the implementation proves to be unstable over long time-scales. While we do not know for certain, this is likely because the source term is unbounded and results in a stiff ODE as is claimed in the analysis. We suggest that the hidden-variable approach, when implemented using a stable formulation, has the ability to surpass the other static sub-grid models for the simple reason that including a transport equation for the sub-grid physics is a more general approach than assuming a steady-state closure. This is analogous to whether one chooses to transport the pressure in a set of fluid equations governing the density and momentum, or assume that the pressure is provided by a closure relation such as the ideal gas law. The former is a more general version of the latter.

\section{Problem Statement -- Landau damping of excited plasma wavepackets}\label{sec2}%- Kinetic Effects in Fundamental Plasma Physics}
Plasmas are systems where an ionized gas exhibits collective behavior through electrostatic or electromagnetic fields. These fields are capable of sustaining oscillations and waves which leads to a discipline rich in non-linear kinetic behavior. The kinetic behavior arises due to wave-particle interactions, where particles moving at the phase velocity of the wave can exchange energy with the wave. One important example of this is Landau damping \cite{Landau1946}, where an excited plasma wave will damp even in the absence of dissipative forces. This inherently requires a kinetic description to describe the physics of the wave-particle interaction driven damping mechanism. The linear version of Landau damping can be included via an analytic closure relation, which we do here. We extend that implementation to include the nonlinear version of this effect in a fluid model through the introduction of a ``hidden variable'' representing the trapped electron population. We start by describing the simple one dimensional fluid model.
\subsection{Fluid model}
The relevant equations are the simplest form of the moment equations for electron plasma transport that are necessary for illustrating our method. The two-moment equations are given by
\begin{align}
    \partial_t n + \partial_x (u~n) &= 0, \label{eq:con} \\
    m n \left(\partial_t u + u ~ \partial_x u\right) &= -\partial_x p + q n E, \label{eq:mom} 
%    p &= \gamma ~ n ~ k_B T, \label{eq:ideal}
\end{align}
where $n=n(t,x), u=u(t,x), p=p(t,x)$ are the density, velocity, and pressure, respectively. We use an adiabatic equation of state as a closure for the equation set,
\begin{equation}
    p = p_0\left(\frac{n}{n_0}\right)^\gamma\;,
\end{equation}
where $\gamma = (f+2)/f$ is the adiabatic index, with $f$ the number of degrees of freedom, which is set to 3.0 for one-dimensional problems. $T$ is the electron temperature and is assumed to be constant and uniform. $E=E(t,x)$ is the electrostatic field given by solving Poisson's equation, $\partial_x E= (Z n_i - e n) / \epsilon_0$ where $n_i$ is the ion density. Because this problem requires electron timescales, it is satisfactory to assume ions are stationary and form a neutralizing background. 

\subsection{Base Model - Linear Landau Damping}\label{sec:epw}
During linear Landau damping, particles absorb energy from the electric field and are accelerated. Because individual particle trajectories are no longer modeled in the fluid approximation, modeling Landau damping requires a modification to include the microphysics. In ref. \cite{hammett_fluid_1990}, a Fourier space operation is used to provide the microphysics closure. %because accurate modeling requires the implementation of a wavenumber dependent damping rate.

% It has been previously shown that linear Landau damping can be included analytically. 

We include Landau damping similarly by adding a damping term to the momentum equation given by eq. \ref{eq:mom}. The modified momentum equation is given by
\begin{align}
    m n \left(\partial_t u + u ~ \partial_x u\right) &= -\partial_x p + q n E + 2 m n~ \nu_L * u, \label{eq:ld-u}
\end{align}
where $*$ is the convolution operator defined by $[a*b](x) \equiv \int a(x^\prime-x) b(x^\prime) dx^\prime$ and $\nu_L$ is the wavenumber dependent Landau damping rate. Since $\nu_{L}$ can be calculated for a range of wavenumbers using numerical rootfinders, we pre-compute the $\nu_L$ for the computational domain, which serves as a convolutional filter that represents the wave-number dependent damping rate. We verify that this implementation reproduces Landau damping by comparing the results against those from a Vlasov code. We also verify that the analytical linear dispersion relation arising from by eqs. \ref{eq:con} and \ref{eq:ld-u} %, and %\ref{eq:ideal}
recovers the Landau damping rate from the imaginary part of the kinetic dispersion relation.

% Theory and simulations show that this results in a distortion of the distribution away from a Maxwellian in the region of velocity space in phase with the electric field.
% This dynamical system, where the electric field and particles exchange large amounts of energy, is commonly known as non-linear Landau damping because the small parameter, the electric field in this case, is large enough to display noticeable second order effects. 

\subsection{Training - Non-linear Landau damping}\label{sec:nlepw}
When the electric field amplitude is large, particles exchange energy with the electric field, and eventually become trapped in the potential of the wave.  The electric field and particles reach a time-averaged steady state during which Landau damping is suppressed and the electric field is not damped. In a weakly-collisional plasma, Landau damping does not completely disappear, but is reduced significantly, and the amount of reduction is a function of the collision frequency, field amplitude, and the wavenumber of the field. This is a distinctly non-linear and kinetic effect that is not captured by the fluid model.

A model equation that enables the field amplitude dependent suppression of Landau damping can be given by
\begin{align}
    m n \left(\partial_t u + u ~ \partial_x u\right) &= -\partial_x p + q n E + 2 ~ mn~\nu_L(k, |E|^k; \theta) * u, \label{eq:ld-nl-local}
\end{align}
where $\nu_L$ is now a function of the wavenumber dependent field amplitude. An analytical solution for that term is not available, but a representation can be machine learned via tuning its weights and biases $\theta$. However, it is readily seen that this implementation is unable to recover effects that are non-local in space, that is, effects that require nearby information, such as wavepacket etching. The formulation, by introducing a spatiotemporally dependent ``hidden variable'' that we use instead is detailed in the next section.

\subsection{Testing - Motivating non-local physics through wavepacket etching}
In the previous subsections, we have been discussing phenomenon that occurs in a idealized periodic system of a single wavelength. In finite-length wavepackets of $\mathcal{O}(10)$ wavelengths, a kinetic effect occurs where the damping rate becomes increasingly non-local because it depends on past dynamics. The effect on the observable is that the back of the wavepacket is damped faster than the front. It was shown in ref. \cite{fahlen_propagation_2009} that the reason behind this is that resonant electrons are accelerated by the wave and travel faster than the group velocity of the wave. Because the electrons in the back of the wave are no longer trapped, non-linear Landau damping no longer occurs and the back of the wave resumes linear behavior. On the other hand, since the electrons from the back are traveling to the front of the wave, the front of the wave continues to experience non-linear effects. For this reason, the back of the wave damps faster than the front. The Euler equations do not discriminate between the resonant electrons and the bulk fluid. Therefore, they are unable to capture this effect and require further modification.

\section{Method}\label{sec3}
\subsection{A spatiotemporally non-local model of Landau Damping}
We modify eq. \ref{eq:ld-nl-local} by adding an auxiliary ``hidden variable'' $\delta=\delta(t,x)$ that represents the resonant electrons. From this section on we write the equations in normalized form using electrostatic units where $v\rightarrow v / v_{th}$, ${x} \rightarrow x/\lambda_D$, ${t} \rightarrow \omega_p t$, ${m} \rightarrow m/m_e$, ${q} \rightarrow q / e$, and ${E} \rightarrow e E / m_e v_{th} \omega_p$ \cite{joglekar_unsupervised_2022}. Here, $v_{th}$ is the thermal velocity, $\omega_p$ is the plasma frequency, and $m_e$, and $e$, are the electron mass, and charge, respectively. Since we seek to minimize Landau damping in their presence, we assume a sigmoidal dependence, similar to that in ref. \cite{tran_fluid_2020}, and the resulting equation is given by
\begin{align}
    m n \left(\partial_t u + u ~ \partial_x u\right) &= -\partial_x p + qnE + 2 ~ mn ~ \frac{\nu_L * u}{1 + \delta^2}. \label{eq:ld-nl-nl}
\end{align}
The growth and transport of $\delta$ is given by
\begin{align}
    \partial_t \delta &= v_{ph} \partial_x \delta + \nu_g ~ \frac{|E * \nu_L|}{1+\delta^2}, \\
    \nu_g &= \nu_g(k, \nu_{ee}, |\hat{E}^k|; \theta_g). \label{eq:deltag}
\end{align}
This model makes use of the knowledge that, since $\delta$ represents the electrons in phase with the wave, $\delta$ must be transported at the phase velocity. It also uses a source term inspired by linearizing the Vlasov-Boltzmann equation. Because $\delta$ is initialized at zero, the proportionality factor of the source term is what is approximated via a machine learned function.

\subsection{\textsc{ADEPT} - Automatic-Differentiation-Enabled Plasma Transport}
To train the model in eq. \ref{eq:deltag}, we need a simulator that can model the set of equations given by
\begin{align}
    \partial_t n + \partial_x (u~n) &= 0,  \label{eq:con2} \\
    m n \left(\partial_t u + u ~ \partial_x u\right) &= -\partial_x p + q nE + 2 ~ mn ~ \frac{\nu_L * u}{1 + \delta^2}. \label{eq:mom2} \\
    %p &= \gamma ~ n ~ k_B T, \\
    p &= n^\gamma, \\
    %\partial_x E &= \frac{Z n_i - e n}{m \epsilon_0}, \\
    \partial_x E &= \frac{q}{m}(n-1), \\
    \partial_t \delta_k &= v_{ph} ~ \partial_x \delta +  \nu_g ~ \frac{|E * \nu_L|}{1+\delta^2}, \label{eq:delta2} \\
     % \nu_g ~ \frac{|E(x) * \nu_L(x-x')|}{1+\delta^2}, \\ %- \nu_d ~ \delta,
    \nu_g &= 10^{f(\tilde{E}_k, \tilde\nu_{ee}, \tilde k; \theta_g)}, \label{eq:deltag2} \\
    f(E, \nu_{ee}, k; \theta) &= 3\tanh\left(g(\tilde \nu_{ee}, \tilde k, |\hat{\tilde{E}}^k|; \theta)\right), \label{eq:deltag3}
\end{align}
where the final term in eq. \ref{eq:mom2} represents the non-local kinetic effect that is added to the typical Euler equations. The closure relation for that term is given by eqs. \ref{eq:delta2}, \ref{eq:deltag2}, \ref{eq:deltag3} where $\nu_g$ is the key coefficient that must be learned from data. 

\subsubsection{Neural Network Details and Input Normalization}
$g(\tilde \nu_{ee}, \tilde k, |\tilde{E}^k|; \theta)$ in eq. \ref{eq:deltag3} is the function representing the output from a feedforward neural network with a depth of 3 and width of 8 units. The intermediate activation function is the hyperbolic tangent function. To ensure that the growth rate does not result in numerical instability, we bound the growth rate so that $10^{-3} < \nu_g < 10^{3}$ by leveraging the properties of the hyperbolic tangent. 

$\tilde{\nu}_{ee} = \left(\log_{10}(\nu_{ee}) + 7\right)/4$ is the normalized electron-electron collision frequency, $\tilde{k} = (k - 0.26) / 0.14$ is the normalized wavenumber, $|\hat{\tilde{E}}^k| = (\log_{10}(|\hat{E}^k| + 10^{-10}) + 10 ) / -10$ is the normalized value of the electric field amplitude of the $k$th mode number. We choose this formulation for normalizing the electric field to avoid possible problems with taking the logarithm of zero. 

\subsubsection{Solver and Simulations} We solve this set using a pseudo-spectral, finite-volume method with a fourth order time-stepper \cite{tsitouras_rungekutta_2011} in a spatially periodic domain. The gradients are computed using Fourier transforms. Similarly, Poisson's equation is also solved spectrally. All the other operations, e.g. advection, are performed in real space.

\section{Results} \label{sec4}
\subsection{Training - Suppression of Landau Damping in large amplitude plasma waves}
\subsubsection{Dataset from kinetic simulations}
We train our model against simulations collected from a Vlasov-Boltzmann equation solver. The equations solved by the fully-kinetic solver are 
\begin{align}
    \partial_t f + v ~ \partial_x f + E~  \partial_v f &= \nu_{ee} ~ \partial_v (vf + v_0^2 ~ \partial_v f), \\
    \partial_x E &= Z n_i - n_e,
\end{align}
where $f=f(t, x, v)$ is the electron distribution function, $Z=1$ is the atomic charge, and $n_i = 1$ is the ion density. Details for this solver are given in refs. \cite{joglekar_vlapy_2020, joglekar_unsupervised_2022}

The fixed parameters are $t_{max} = 500 \omega_p^{-1}, \Delta t = 0.5, N_x = 64, N_v = 2048$. In all simulations, we simulate a single mode wave. The parameters of the temporal envelope of the external forcing term are $t_w = 40 \omega_p^{-1}, t_r = 5 \omega_p^{-1}, t_c = 40 \omega_p^{-1}$. The implementation of the forcing term is given in ref. \cite{joglekar_unsupervised_2022}.

\begin{figure}
    \centering
    \includegraphics[width=\textwidth]{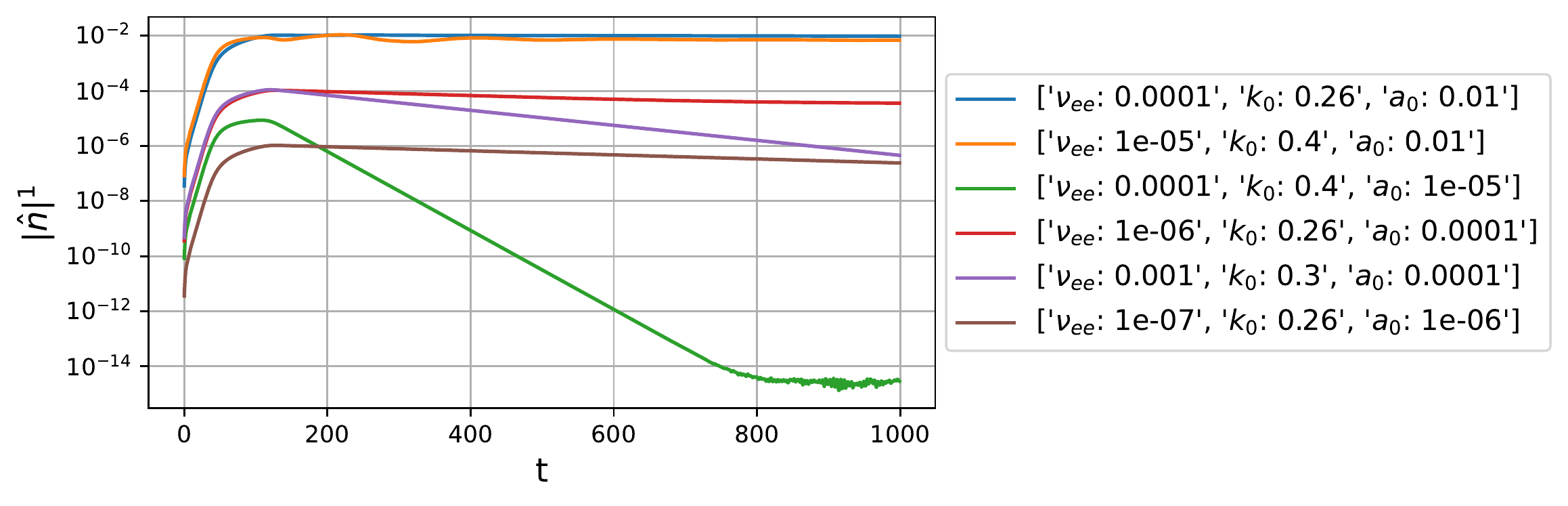}
    \caption{The training data are single mode simulations of plasma waves that are excited by a short duration driver using the fully-kinetic, Vlasov formulation. The training involves reproducing these excitations with the simulations using the modified fluid equations.}
    \label{fig:trainingdata}
\end{figure}

The training dataset comprises 200 first-principles simulations. The parameters that are varied are given by
\begin{align}
    \nu_{ee} \in [10^{-7}, 10^{-6}, 10^{-5}, 10^{-4}, 10^{-3}], \label{eq:nuparams} \\ 
    k \in [0.26, 0.28, 0.3, 0.32, 0.34, 0.36, 0.38, 0.4], \label{eq:kparams}\\ 
    a_0 \in [10^{-6}, 10^{-5}, 10^{-4}, 10^{-3}, 10^{-2} ].\label{eq:aparams}
\end{align}
A few examples are shown in fig. \ref{fig:trainingdata}.

\subsubsection{Fluid Modeling}
The fluid simulations used for training the model have different spatial and temporal resolutions, given by $N_x = 16, \Delta t = 0.025$, but are performed over the same spatial and temporal domains as the kinetic simulations.

Since we seek to replicate the behavior of a single-mode electron plasma wave with our fluid model, our training objective function is to minimize the mean squared error between the timeseries from the Vlasov and fluid simulations. Formally, the loss function is given by
\begin{align}
    \vec{p} &= (a_0, \nu_{ee}, k \lambda_D) \\
\mathcal{L} &= \left[\frac{1} {N_t}\sum_{t=250 \omega_p^{-1}}^{450 \omega_p^{-1}}\left(\log_{10}|\hat{n}^{1}_{f}|(\vec{p}; \theta) - \log_{10}|\hat{n}^{1}_V(\vec{p})|\right)^2\right], \label{eq:loss}
\end{align}
where $\vec{p}$ is the parameter vector that consists of unique combinations of the parameters listed in eqs. \ref{eq:nuparams}, \ref{eq:kparams}, and \ref{eq:aparams}, $|\hat{n}^{1}|$ is the time history amplitude of the first mode of the density profile with the subscripts of $f$ and $V$ correspond to fluid and Vlasov, respectively. It is important to note that we compare the logarithm of the timeseries in the loss function. This is because the wave amplitude can be damped over many orders of magnitude. Without leveraging the logarithm, the wave amplitudes at the end of the simulation will be negligible in comparison to that near the beginning and would effectively be ignored by the gradient descent algorithm.

\subsubsection{Assessment}
\begin{figure}[h]
    \centering
    \includegraphics[width=0.8\textwidth]{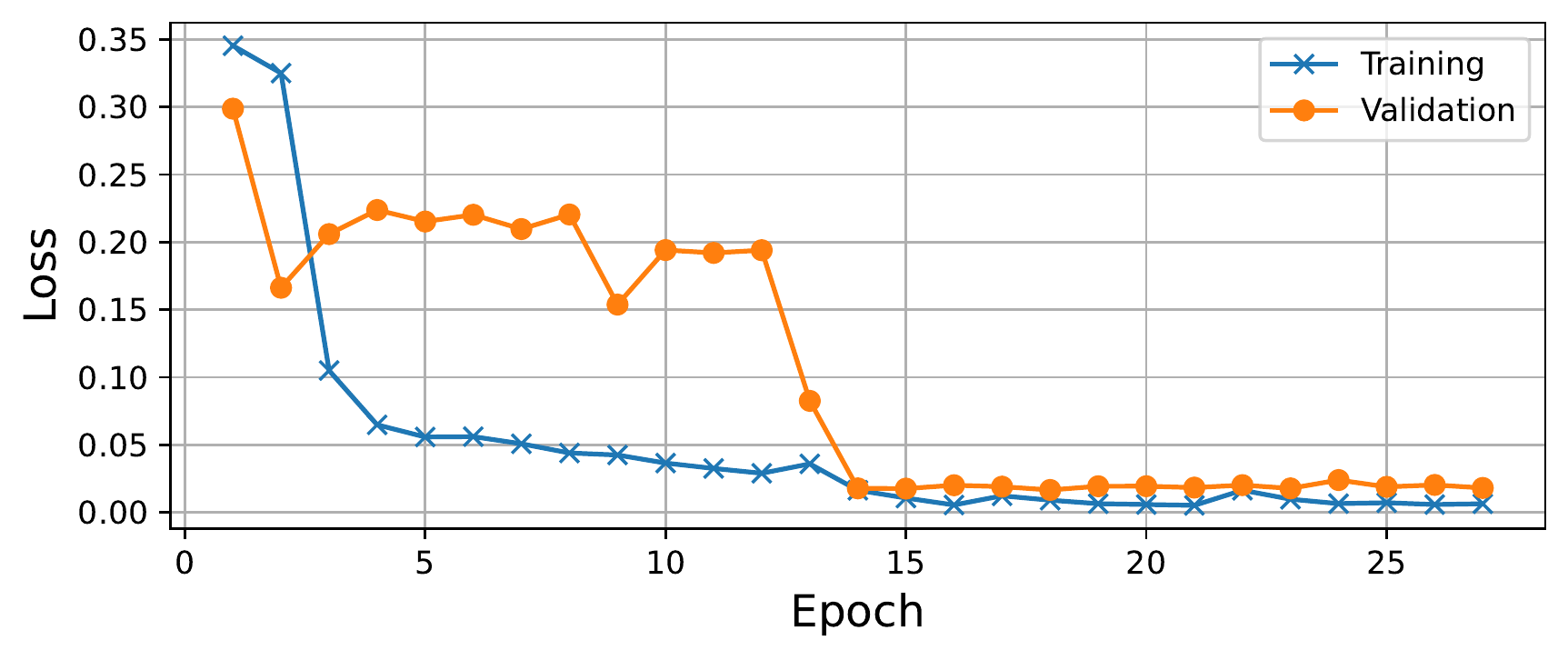}
    \caption{The training and validation loss is plotted against the number of epochs.}
    \label{fig:training}
\end{figure}
We use a 90\%/10\% training and validation split. In fig. \ref{fig:training}, we show the mean squared error over the training epochs for training and validation. Training was performed using A10G GPUs. We implement data parallel training by running $N_\text{batch}$ simulations in parallel and averaging their gradients on the primary node that is running the optimization algorithm. We use ADAM with a learning rate of 0.004 as the optimization algorithm. We stop training after the performance has saturated for many epochs. 

\begin{figure}[ht]
     \centering
     \begin{subfigure}[b]{0.4\textwidth}
         \centering
         \includegraphics[width=\textwidth]{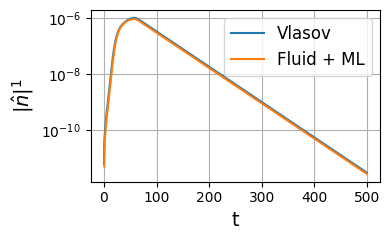}
         \caption{Linear Landau damping}
         \label{fig:train-a}
     \end{subfigure}
     \hfill
     \begin{subfigure}[b]{0.55\textwidth}
         \centering
         \includegraphics[width=\textwidth]{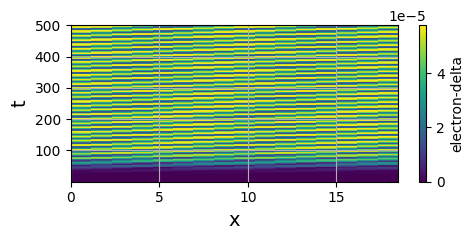}
         \caption{$\delta(t, x)$}
         \label{fig:train-b}
     \end{subfigure}
     \\
          \begin{subfigure}[b]{0.4\textwidth}
         \centering
         \includegraphics[width=\textwidth]{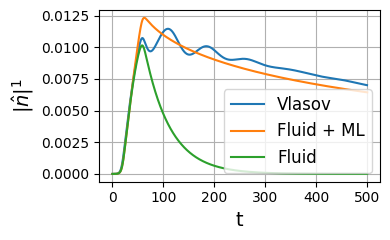}
         \caption{Non-linear Landau damping}
         \label{fig:train-c}
     \end{subfigure}
     \hfill
     \begin{subfigure}[b]{0.55\textwidth}
         \centering
         \includegraphics[width=\textwidth]{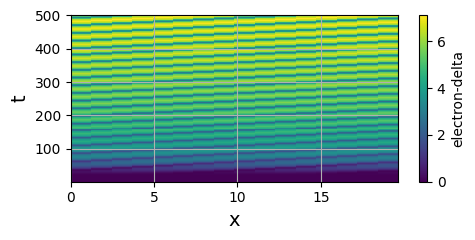}
         \caption{$\delta(t, x)$}
         \label{fig:train-d}
     \end{subfigure}
    \caption{(a) A small amplitude simulation. In this simulation, $\delta$ is not needed to modify the damping rate because the system exhibits linear dynamics. Accordingly, (b) shows that $\delta$ remains small. (c) A large amplitude simulation with non-linear Landau damping. $\delta$ is needed to modify the damping rate. The linear fluid model is incapable of reproducing the desired physics. To recover the non-linear effects, the simulation leverages $\delta$ and (d) shows that $\delta \approx 7$.}
    \label{fig:trainingsims}
\end{figure}

Figure \ref{fig:trainingsims} shows the result of two sample simulations. We choose to show the result of these two sets of parameters because they are representative examples of two extremes that the machine learning model must be able to handle.

In fig. \ref{fig:trainingsims}(a), we show the results from a simulation with $\nu_{ee} = , k_0 = 0.34, a_0 = 10^{-6}$. The two simulations agree well. This is a parameter regime where linear Landau damping is expected, and therefore, the theory from sec. \ref{sec:epw} applies. Therefore, theory dictates that to reproduce the desired behavior, $\delta$ must remain small. Figure \ref{fig:trainingsims}(b) shows that $\delta \approx 10^{-5}$ as the theory suggests it should. Because we train using the logarithm, as stated in eq. \ref{eq:loss}, we are able to capture the damping over five orders of magnitude.

In fig. \ref{fig:trainingsims}(c), we show the results from a simulation with $\nu_{ee} = 10^{-5}, k_0 = 0.32, a_0 = 10^{-2}$. The two simulations agree well. In this parameter regime, kinetic theory dictates that there is a complex interplay between the collisions, represented by $\nu_{ee}$, and the field amplitude, represented by $a_0$. Here we expect that to reproduce the desired behavior, $\delta$ must be some finite quantity that appreciably, but not completely suppresses the damping. Figure \ref{fig:trainingsims}(d) shows that $\delta \approx 7$ and the damping is suppressed by a factor of 50.

%We also show the best and worst performers in the training and validation sets ignoring the cases where linear Landau damping is expected because we consider those to be uninteresting with respect to the goal of this work, to reproduce non-linear Landau damping. If $\delta = 0$, linear Landau damping is recovered and the machine learning model does not play an active role in the dynamics. This is as expected for simulations with a small $a_0$, large $k$ and $\nu_{ee}$. However, it is worth noting that the model does not anomalously generate $\delta$ in these cases, and does recover the expected result. 

%The best performance on the training set occurs on a simulation where Landau damping is effectively fully suppressed. The model is responsible for generating a delta large enough to suppress Landau damping. The worst performance on the training set occurs on a simulation with moderate suppression of Landau damping. However, we deem this to be a success because before this, there did not exist a way to capture this moderate suppression at all. This simply suggests that further training and refinement of the model can improve the performance on these moderate cases. We find that the validation set performance is similar which suggests the model is able to perform interpolation or relatively nearby extrapolation with the parameters of the data reasonably well.

Now that we have been able to learn a microphysics model, we use the coupled set of PDEs along with the newly acquired model weights to simulate a novel geometry and verify its performance qualitatively. 

\subsection{Testing - Extending to unseen geometries and modeling unseen phenomena}

\begin{figure}[ht]
     \centering
     \begin{subfigure}[b]{0.495\textwidth}
         \centering
         \includegraphics[width=\textwidth]{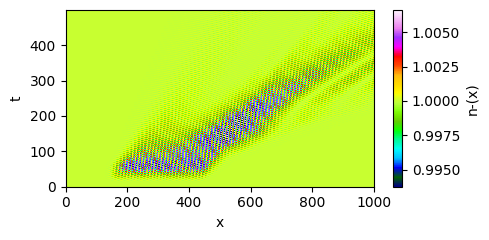}
         \caption{Vlasov}
         \label{fig:nl-wp-n-v}
     \end{subfigure}
     \begin{subfigure}[b]{0.495\textwidth}
         \centering
         \includegraphics[width=\textwidth]{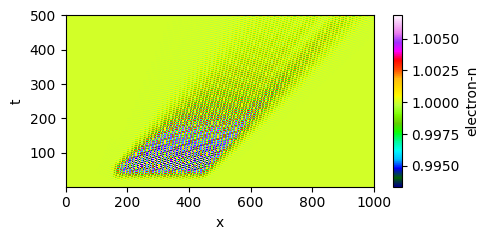}
         \caption{Fluid + Local Model}
         \label{fig:nll-wp-n-f}
     \end{subfigure}
     % \begin{subfigure}[b]{0.495\textwidth}
     %     \centering
     %     \includegraphics[width=\textwidth]{l-wp-n-v.png}
     %     \caption{Vlasov - $n(t,x)$}
     %     \label{fig:l-wp-n-v}
     % \end{subfigure}
     \\
     \begin{subfigure}[b]{0.495\textwidth}
         \centering
         \includegraphics[width=\textwidth]{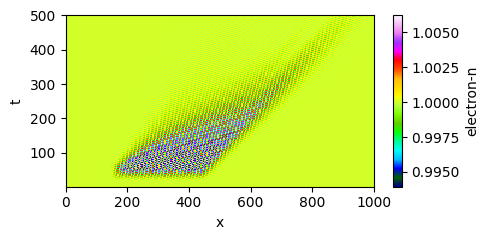}
         \caption{Fluid + $\delta(t, x)$}
         \label{fig:nl-wp-n-f}
     \end{subfigure}
     \begin{subfigure}[b]{0.495\textwidth}
         \centering
         \includegraphics[width=\textwidth]{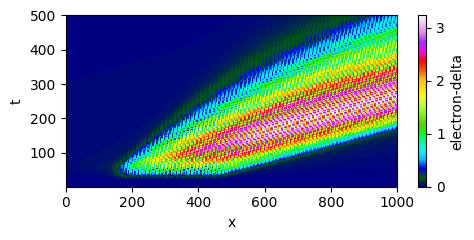}
         \caption{$\delta(t,x)$}
         \label{fig:nl-wp-delta}
     \end{subfigure}
     % \begin{subfigure}[b]{0.495\textwidth}
     %     \centering
     %     \includegraphics[width=\textwidth]{l-wp-n-f.png}
     %     \caption{Fluid + ML $n(t, x)$}
     %     \label{fig:l-wp-n-f}
     % \end{subfigure}
     % \\
     
     % \begin{subfigure}[b]{0.495\textwidth}
     %     \centering
     %     \includegraphics[width=\textwidth]{l-wp-delta.png}
     %     \caption{$\delta(t,x)$}
     %     \label{fig:l-wp-delta}
     % \end{subfigure}
    \caption{(a) shows the ground-truth density profile over space and time for a large amplitude excitation. (b) shows that using a local model for Landau damping is insufficient in modeling this excitation. (c) shows the result of a simulation that uses machine-learned $\delta(t,x)$ model which shows better agreement and reproduction of the non-local damping. (d) shows that $\delta$ grows to $3$ in the case of this large amplitude perturbation to account for the non-locality. It is important to note that the scale length over which the model is trained in fig. \ref{fig:trainingsims} is significantly different than the scale length to which the model is applied here.}
    \label{fig:testsims}
\end{figure}

In the previous section, we trained on a dataset of systems where a single wavelength wave was excited in a box of one wavelength in length.  To assess the performance of our learned model in unseen geometries, we perform simulations where the box size is increased by a factor of 100 and a 20 wavelength long, finite-length wave is modeled. This tests the ability of the machine learned model combined with the hidden variable approach that was trained on a limited geometry to generalize to larger length scales and different boundary conditions. To verify the fidelity, we compare the results of the learned fluid simulations to their kinetic counterparts acquired by using the Vlasov-Boltzmann solver.

In refs. \cite{fahlen_propagation_2009, joglekar_unsupervised_2022}, it has been shown that finite-length wavepackets damp non-uniformly due to a non-linear, kinetic phenomenon. In fig. \ref{fig:testsims}(a), we show an example of this phenomenon. Figure \ref{fig:testsims}(b) shows that if a local implementation of Landau damping is used, for example, like the one given by eq. \ref{eq:ld-nl-local}, the non-uniform damping will not be reproduced because the damping rate will be the same along the length of the wavepacket. In contrast, fig. \ref{fig:testsims}(c) shows that the implementation using the machine-learned dynamical hidden variable approach is much more capable of reproducing the results from the ab-initio simulation. Figure \ref{fig:testsims}(d) confirms that $\delta$ grows to account for the suppressed damping, and then is transported to account for the non-locality in space and time. 

\begin{figure}[h]
     \centering
     \begin{subfigure}[b]{0.495\textwidth}
         \centering
         \includegraphics[width=\textwidth]{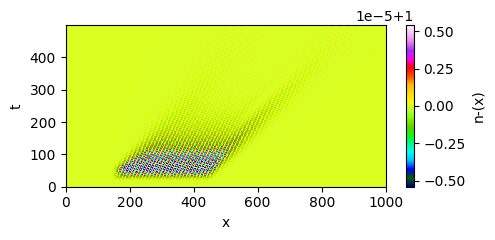}
         \caption{Vlasov}
         \label{fig:l-wp-n-v}
     \end{subfigure}
     \begin{subfigure}[b]{0.495\textwidth}
         \centering
         \includegraphics[width=\textwidth]{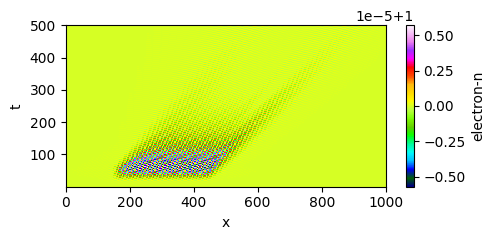}
         \caption{Fluid + Local}
         \label{fig:l-wp-n-f-local}
     \end{subfigure}
     \\
     \begin{subfigure}[b]{0.495\textwidth}
         \centering
         \includegraphics[width=\textwidth]{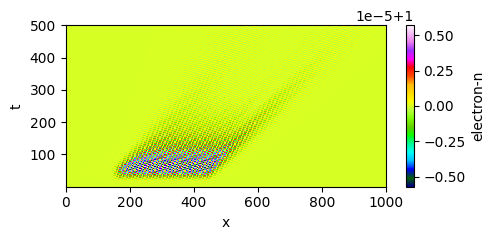}
         \caption{Fluid + $\delta(t,x)$}
         \label{fig:l-wp-n-f}
     \end{subfigure}
     \begin{subfigure}[b]{0.495\textwidth}
         \centering
         \includegraphics[width=\textwidth]{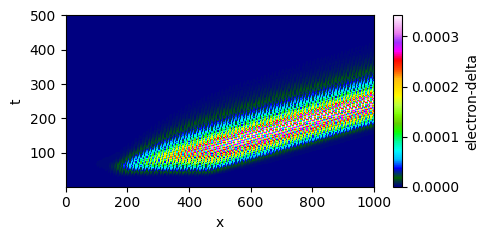}
         \caption{$\delta(t,x)$}
         \label{fig:l-wp-delta}
     \end{subfigure}
    \caption{(a) shows the ground-truth density profile over space and time for a small amplitude excitation. In contrast to fig. \ref{fig:testsims}, (b) and (c) agree well here because at small amplitudes, the damping remains local and the non-local model is expected to converge to the local result. Correspondingly, (d) shows that $\delta$ remains small.}
    \label{fig:testsims-l}
\end{figure}

Similar to that in fig. \ref{fig:trainingsims}, we also show the behavior in the linear regime exemplified by a small amplitude simulation in fig. \ref{fig:testsims-l}(a).  In this regime, it is expected that Landau damping remains local as shown in fig. \ref{fig:testsims-l}(b). \ref{fig:testsims-l}(d) shows that the model does not anomalously generate large $\delta$. Because of this, in fig. \ref{fig:testsims}(c), we see that linear Landau damping proceeds as expected and the non-local model recovers the local result.

In this work, we seek to highlight the fact that because we choose to embed the machine learning model into a set of coupled PDEs that are capable of generalizing, we can train the model on a simple system and can scale the learned physics to a significantly more complex domain. In this case, we scale a single-wavelength periodic system to a finite-length wavepacket in an open boundary simulation but this model could be very well be extended to multiple spatial dimensions. If the use-case were to acquire as accurate a model as possible in order to perform multi-scale modeling of an actual experiment, one could and should train the model on as many geometries as is feasible. 

\section{Conclusion}\label{sec5}
The marriage of numerical PDE solvers and deep learning driven models is more than coincidental because they are both products of numerical computing. The advent of scalable and performant scientific computing libraries that are capable of automatic differentiation, e.g. JAX and Julia, has enabled a seamless interchange of ideas between the two paradigms where the lines between each continue to blur. 

In this work, we show how the two paradigms can be intimately woven together and leveraged towards performing multiscale modeling where the equation set does not natively contain all the necessary microphysics, and where the microphysics model itself has proven to be challenging to construct using conventional methods. By training against a dataset with a relatively simple geometry and parameters, we are able to learn a model that can emulate spatio-temporally non-local physics over broad parameter ranges and geometries, physics which can otherwise only be described by computationally expensive higher-dimensional PDE solvers. The key to preserving the spatio-temporally non-local dynamics is providing a dynamical system for a hidden variable that governs the physics. By allowing for growth and transport of the hidden variable, we enable non-locality in time and space. 

We envision that a model that leverages machine-learned hidden variables can be used for various phenomenon inside and outside of plasma physics. For example, hot electrons created from laser-plasma instabilities are deleterious towards successful fusion experiments. Accurate modeling of their dynamics and energetics in fully integrated hydrodynamics simulations of fusion experiments can help substantially improve the predictive power of simulations and improve the results of fusion experiments.

%To provide a plasma physics example, the problem of accurately modeling non-local heat flow has long plagued fusion experiments and many analytically derived models have been proposed. While there has been some preliminary work in modeling non-local heat flow using neural networks \cite{lamy_modeling_2022}, a method by which to acquire a model that is flexible to different discretizations and geometries has yet to be published. We believe the method proposed here could be a path forward. 

\section{Data Availability Statement}
The data that support the findings of this study are openly available at the following URL:
\url{https://github.com/ergodicio/adept}

\section{Conflict of Interest}
Authors declare no competing interests.

\section{Acknowledgments}
We acknowledge funding from the DOE Grant \# DE-SC0016804. 

We use and would like to acknowledge open-source software from the Python ecosystem for scientific computing. Specifically, we use NumPy \cite{Harris2020}, SciPy \cite{scipy_10_contributors_scipy_2020}, matplotlib \cite{Hunter:2007}, Xarray \cite{Hoyer2017}, JAX \cite{bradbury_jax_2018}, and Diffrax \cite{kidger2021on} while the neural network is managed using Equinox \cite{kidger2021equinox}. The numerical experiments and data are managed using MLFlow \cite{10.1145/3399579.3399867} as discussed in ref. \cite{feister_control_2023}.

A. J. also thanks D. J. Strozzi and W. B. Mori for useful discussions.

\section*{References}
\bibliographystyle{iopart-num}
\bibliography{bib}

\end{document}